\documentclass[aps,prl,twocolumn,preprintnumbers,groupedaddress,nofootinbib]{revtex4}

\pdfoutput=1

\usepackage{graphicx}
\usepackage{amsfonts}
\usepackage{amssymb}
\usepackage{amsmath}
\usepackage{slashed}
\usepackage{hyperref}
\usepackage{url}
\usepackage{multirow}
\begin{document}
\include{declare}

\title{Don't Miss the Displaced Higgs at the LHC Again}

\author{Matthew R.~Buckley$^{1}$, Valerie Halyo$^2$, and Paul Lujan$^2$}
\affiliation{$^1$ Department of Physics and Astronomy, Rutgers University, Piscataway, USA}
\affiliation{$^2$ Department of Physics, Princeton University, Princeton, USA}

\preprint{}
\date{\today}

\begin{abstract}
A signature often found in non-minimal Higgs sectors is Higgs decay to a new gauge-singlet scalar, followed by decays of the singlets into Standard Model  fermions through small mixing angles. The scalar decay can naturally be displaced from the primary vertex. The present experimental constraints on such models are extremely weak, due to low (or zero) trigger rates for the resulting low $p_T$ displaced jets. In this letter, we highlight the advantages of integrating into the trigger system massively parallel computing and coprocessors based on Graphics Processing Units (GPUs) or the Many Integrated Core (MIC) architecture. In particular, if such coprocessors are added to the LHC experiments' high level trigger systems, a fast Hough transform--based trigger performed on this hardware would result in significant improvement to displaced searches, sufficient to discover long-lived Higgs models with a small amount of luminosity in Run II at the 14~TeV LHC.
\end{abstract}
\maketitle

The discovery of the 125~GeV Higgs by ATLAS and CMS \cite{Aad:2012tfa,Chatrchyan:2012ufa} has been a triumph for both experimental and theoretical physics \cite{Englert:1964et}. As we move into the Higgs Era, there is huge potential for new physics discovery from precision measurements of the Higgs' branching ratios (BRs) into Standard Model (SM) or non-SM particles and searches for an extended Higgs sector. Though the present measurements show no clear signal of deviation from the SM predictions \cite{Aad:2013wqa}, many channels remain unexplored, and the story is far from complete (see {\it e.g.}~Ref.~\cite{Curtin:2013fra}). In this letter, we will show that the experimental triggers currently in use at CMS and ATLAS are poorly suited to discover extensions of the Higgs sector involving long-lived particles. We demonstrate that such scenarios could be easily discovered in a small amount of LHC data with the addition of new massively parallel trigger algorithms, preferably with accelerators such as Graphics Processing Units (GPUs) or Xeon Phi coprocessors integrated into the trigger system.

In a broad class of SM Higgs sector extensions, the Higgs can develop decay modes which terminate in vertices displaced a macroscopic distance from the beam. The simplest example contains just a single extra particle: a gauge singlet real scalar $X$ \cite{Strassler:2006ri}. Assuming just a single $SU(2)_L$ Higgs doublet $H$ (with physical Higgs $h$ after electroweak symmetry breaking), this model has the potential 
\begin{eqnarray}
V(H,X) & = & - \mu^2 H^2+\lambda H^4+M^2 X^2 + \lambda_X X^4  \\
& & + \lambda_{HX} H^2X^2+aX+bX^3+cXH^2. \nonumber
\end{eqnarray}
The scalar $X$ has a tree-level coupling $\lambda_{HX}$ to the Higgs, resulting in $\Gamma(h \to XX) \sim (10~\mbox{GeV})\lambda_{HX}^2$ after electroweak symmetry breaking (assuming $m_X \ll m_h$). This typically results in large $h\to XX$ BRs when the $X$ is kinematically accessible, though the experimental upper limit on the total Higgs width \cite{CMS:2014ala} sets a bound on $\lambda_{XH} \lesssim 0.04$. If $a=b=c=0$, then the potential has a $\mathbb{Z}_2$ symmetry and the $X$ is stable. Breaking this symmetry allows for the $X$ to have small mixing with the Higgs, {\it i.e.}~the mass eigenstate is $X+\epsilon h$. The small $\epsilon \ll 1$ parameter is technically natural, as $\epsilon = 0$ is a point of enhanced symmetry. The $X$ can only decay through this admixture with the Higgs, so the couplings of the $X$ to SM fermions are proportional to the fermion masses. 

As a result, we expect the $X$ to decay to the most massive fermions kinematically available ($W$ and $Z$ decays being highly suppressed, as we assume $m_X < m_h/2$). The width to $b$-quarks is $\Gamma(X\to b\bar{b}) \sim (\epsilon^2/16\pi)m_b^2 m_X/v^2$ (assuming $m_b \ll m_X$). This easily accommodates long lifetimes for the $X$, with $\Gamma^{-1} = c\tau \sim (10^{-5}/\epsilon)^{2}~\mbox{mm}$. 

This simple model can be easily generalized to an extended Higgs sector with multiple doublet Higgses. Each Higgs can have different BRs into a pair of $X$ particles, followed by displaced decays of $X$ into SM fermions, with the ratios set by the details of the particular model. Similar phenomenology can be found in hidden valley \cite{Strassler:2006im} or supersymmetric models \cite{Graham:2012th}.

Much of the theoretical and experimental effort has concentrated on the prompt decays of the 125 GeV Higgs. These searches are not very sensitive to possible displaced Higgs decay modes. Though CMS and ATLAS have performed searches for displaced decays \cite{Aad:2013txa,Chatrchyan:2012cg,CMS:2013oea}, they place only weak constraints on Higgs displaced decays, as the low masses involved make it difficult to pass jet- or MET-based triggers at the 7 or 8 TeV LHC. The situation at the 14~TeV LHC will only be more difficult, as the trigger thresholds will be raised. Constraints have also been set by D\O\ \cite{Abazov:2009ik} and CDF \cite{Aaltonen:2011rja}. Strong general bounds are placed by the invisible width of the Higgs \cite{Bai:2011wz,Belanger:2013kya}. The global Higgs fits limit the invisible Higgs BR to $<23\%$ at $2\sigma$, assuming all other couplings are SM-like, and $<60\%$ if both the tree- and loop-level couplings are allowed to deviate from SM expectations \cite{Belanger:2013kya}. Direct bounds can be placed on heavier Higgses decaying with displaced vertices, as the larger masses increase the trigger efficiencies, but much of the parameter space is inaccessible \cite{Chatrchyan:2012cg}. 

The search for physics with displaced decays can significantly improved by reconfiguring the high level triggers (HLTs) to use  massively parallel computing (MPC) architecture and enhancing the computer farm with accelerators such as GPUs or Xeon Phi. This cutting edge technology provides not only the means to speed up calculations for existing triggers, but also the opportunity to develop new complex algorithms that select events that previously would have evaded detection. Such accelerators can perform track reconstruction based on Hough transformations \cite{hough} in a few milliseconds at luminosity conditions that yield 250 pileup interactions per event  (${\cal L} \sim 10^{35}~\mbox{cm$^{-2}$s$^{-1}$}$).  

We should note that this letter does not mean to imply that machine vision and pattern recognition algorithms are uniformly a more appropriate trigger solution compared to the conventional Combinatorial Track Finder (CTF) algorithms that are currently used in high energy collider physics. Rather, these new techniques can provide a complementary method of identifying interesting events, first studied by the authors~\cite{Halyo:2013iba} due to the similarities between the problems inherent to triggering on events with unique topological objects and computer image processing analysis. 

The Hough transform  differs from the CTF approach in that it does not operate on localized features of a data set. Rather, the technique is more holistic, operating on the entire pattern of tracking information from the detector as an single image. The use of new triggers based on these transformations would allow selection of events at the trigger level with ``interesting'' structure (such as displaced decays) in the detector activity \cite{Halyo:2013iba} that would be a smoking gun for new physics. The applicability of these image-processing techniques to new physics searches has been considered previously in a LHC search for topological objects such as black holes or displaced jets \cite{Halyo:2013cza}. This letter demonstrates the possible extension of the LHC physics reach in the well-motivated displaced Higgs channels if MPC architecture were integrated into the hardware of the trigger system.

We now consider the search channel $h \to XX$, followed by a displaced $X \to b\bar{b}$ decay, assuming a MPC-based trigger at CMS or ATLAS for the LHC14 run. This channel has been investigated using a simulation of the present CMS trigger menu \cite{Halyo:2013yfa}, for the 125~GeV Higgs decaying to $15-40$~GeV $X$ with a BR of 20\%, where the $X$ then decays with an average displacement of $1-10$~mm. In this study, we consider $X$ decays with larger $\Gamma^{-1} = c\tau$, in the range $10-600$~mm. 

Considering only $X\to b\bar{b}$ decays, simulated events are run through an emulation of the L1 trigger \cite{Brooke:2013hnf}, which requires relatively low calorimeter activity. There are many possible L1 triggers that this signal model can pass: triggers based on single- or double-jets, muon+jet (from semileptonic $b$ decays), $e/\gamma$ (which identify spatially small deposits in the electromagnetic calorimeter), or, at higher $m_H$ values, $H_T$ triggers which consider the total scalar sum of all jet energies in the event. In addition, for $c\tau \gtrsim 300$~mm, the displacement of the jets can cause an apparent transverse energy imbalance in the detector, which provides another alternative trigger path through missing $E_T$ triggers. At lower $c\tau$ values, di-jet triggers are more important. However, we emphasize that the L1 trigger menu evolves over time, as the operating conditions of the LHC change.

\begin{figure}[t]

\includegraphics[width=0.9\columnwidth]{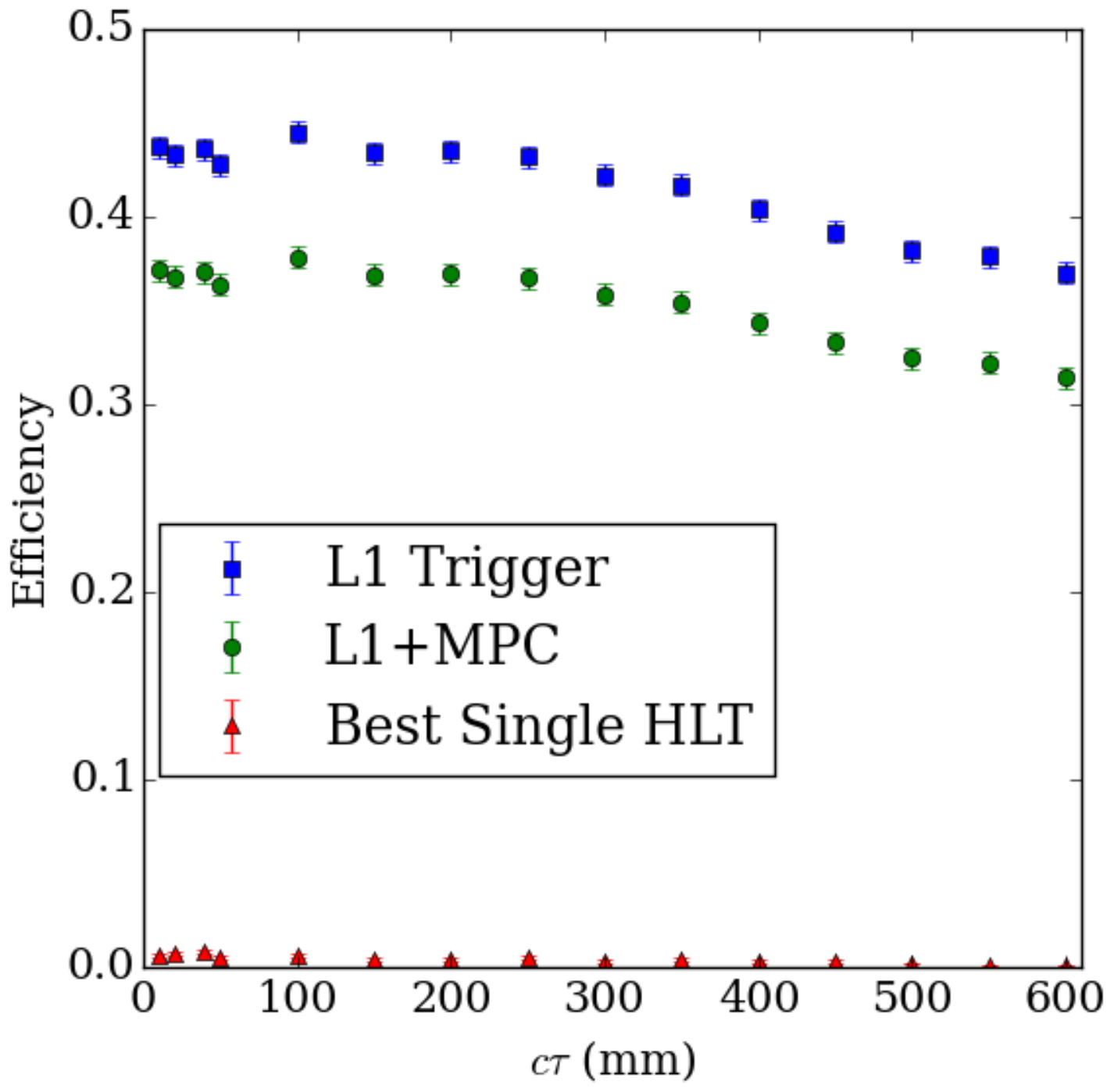}

\caption{Trigger efficiencies for $h\to XX$, $X\to b\bar{b}$ as a function of $c\tau$ with $m_X = 20$~GeV. CMS responses are shown for the full CMS L1 trigger, the single best CMS HLT (a displaced jet trigger, see text), and the MPC trigger applied after the L1. \label{fig:125higgsctautrigger}}
\end{figure}

Events passing the L1 are then passed to the proposed MPC trigger. Approximately 85\% of the events passing the L1 have displaced tracks identified by the Hough transform, independent of $c\tau$ for the values considered. We show in Figure~\ref{fig:125higgsctautrigger} the trigger efficiencies for the currently implemented CMS L1 and the MPC, setting $m_X = 20$~GeV and scanning $c\tau$ from 10 to 600~mm. 

The 2012 8~TeV CMS HLT trigger which has the highest efficiency for events of this class of model is the displaced jet trigger used in Ref.~\cite{CMS:2013oea}, which requires an $H_T$ of at least 250 GeV and identifies displaced jets by requiring a transverse momentum of at least 60 GeV, that most tracks in the jet have an impact parameter greater than 300~$\mu$m, and that no more than $15\%$ of the total jet energy be from tracks with impact parameters less than 500~$\mu$m. To avoid dependence on CMS simulation details, we have constructed a simulation of this trigger relying purely on generator-level information and our own detector simulation. Although this simulation is based on the CMS detector, the situation at ATLAS is expected to be similar. The low efficiency for the current experimental displaced trigger at HLT, as compared to L1 followed by MPC, is primarily due to the high $H_T$ requirement. We show the results of the simulated HLT trigger in Figure~\ref{fig:125higgsctautrigger}.

\begin{figure}[t]

\includegraphics[width=0.9\columnwidth]{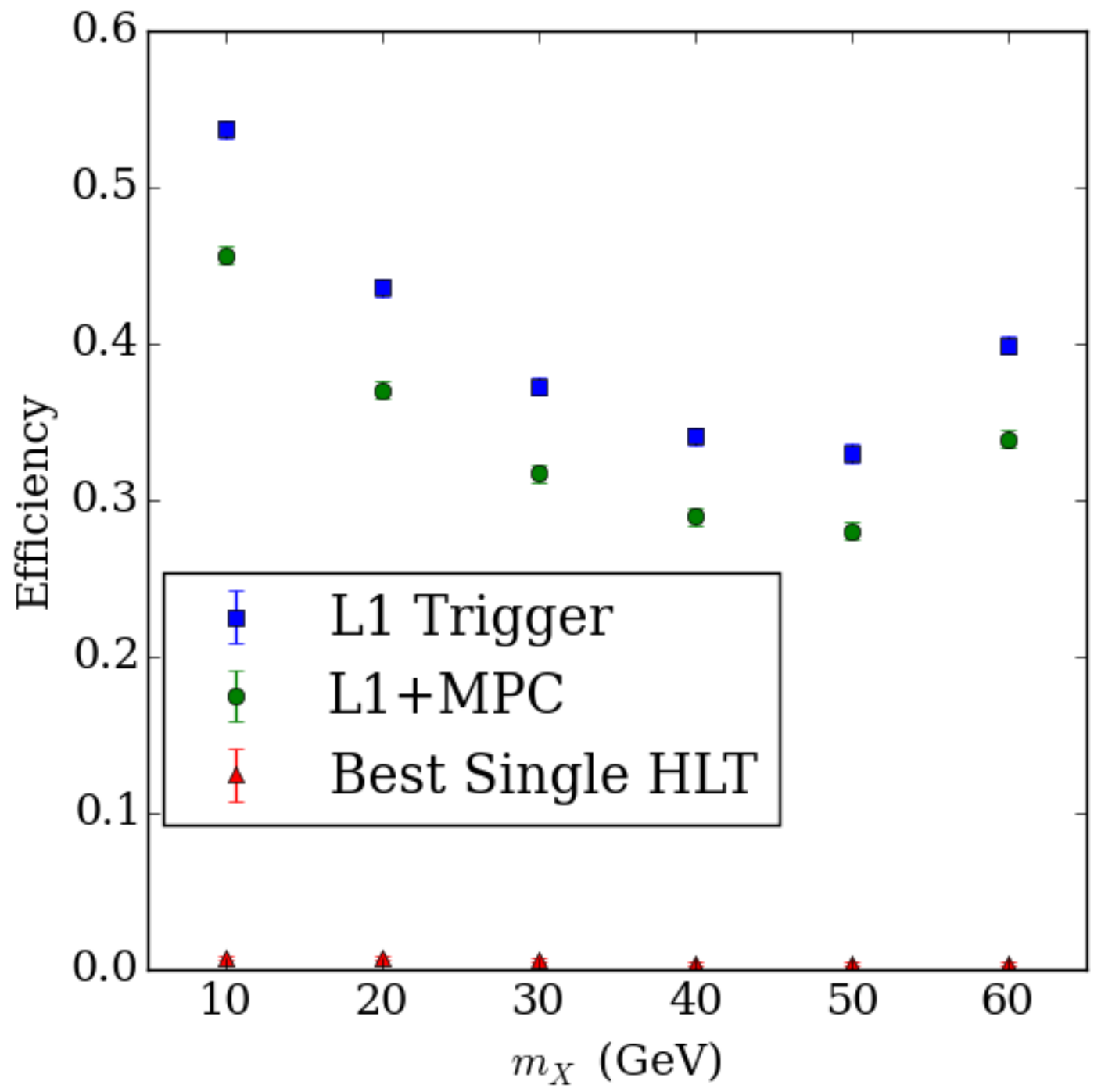}

\caption{Trigger efficiencies for $h\to XX$, $X\to b\bar{b}$ as a function of $m_X$ with $c\tau = 40$~mm. CMS responses are shown for the full CMS L1 trigger, the single best CMS HLT (a displaced jet trigger, see text), and the MPC trigger applied after the L1. \label{fig:125higgsmxtrigger}}
\end{figure}

\begin{figure}[t]

\includegraphics[width=0.9\columnwidth]{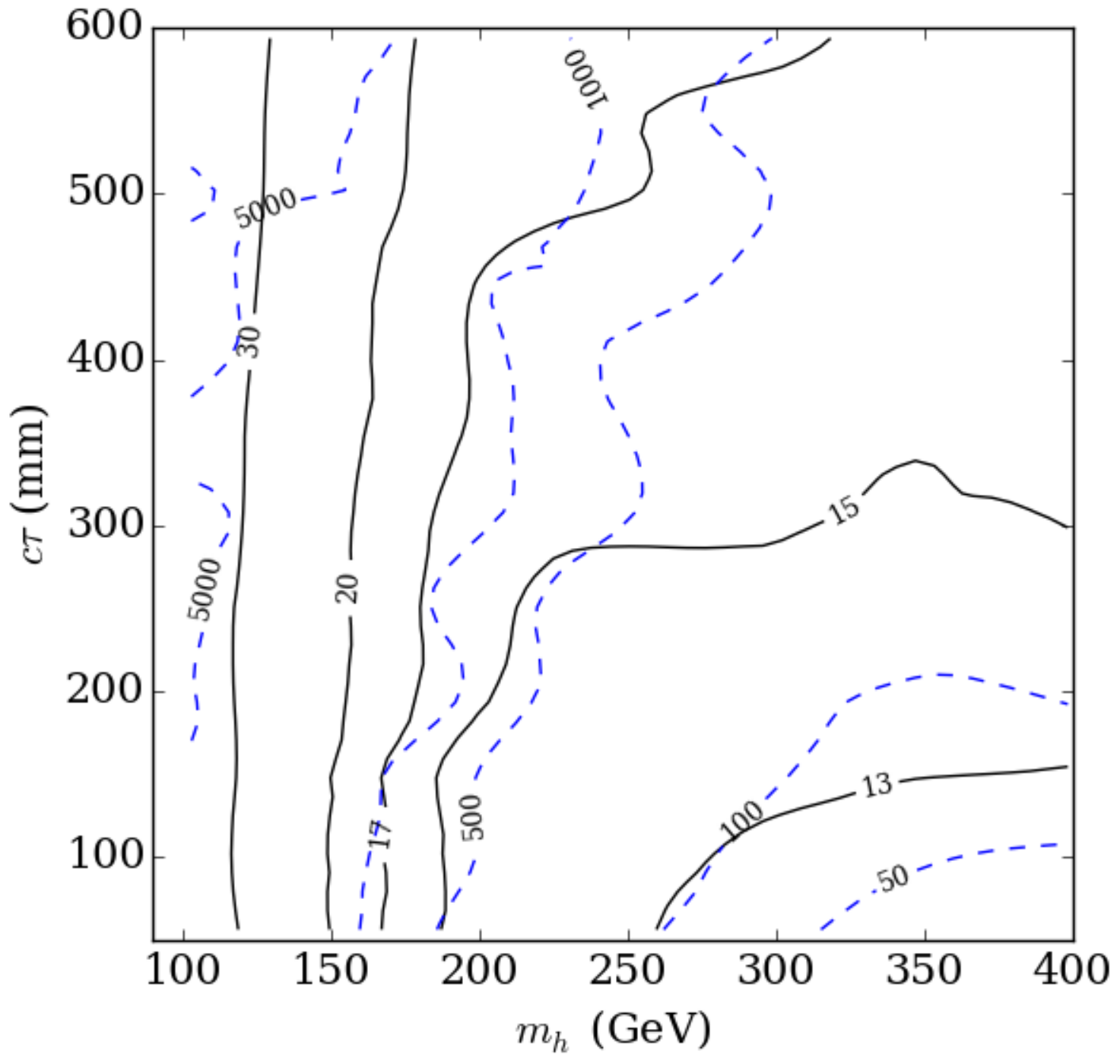}

\caption{Production cross section times BR (in fb) required for 10 event observation in 1~fb$^{-1}$ at the LHC14 for a non-SM $h'$ decaying to pairs of 20~GeV $X$, followed by displaced $X\to b\bar{b}$, as a function of $m_{h'}$ and $c\tau$. Required $\sigma \times$BR are shown for MPC trigger (solid black lines) as well as for the most effective HLT (dashed blue lines). \label{fig:massctau}}
\end{figure}

In Figure~\ref{fig:125higgsmxtrigger}, we again show the trigger efficiencies for the $h \to XX \to 4b$ events, now holding $c\tau =  40$~mm fixed and varying $m_X$. Here, and throughout this paper, signal events were generated using {\tt Pythia6} \cite{Sjostrand:2006za}. 10000 events were generated for each parameter point in our one-dimensional scans over $c\tau$ and $m_X$ (Figures~\ref{fig:125higgsctautrigger} and \ref{fig:125higgsmxtrigger}), and 1000 events were generated per point for our two-dimensional scan (Figure~\ref{fig:massctau}). The uncertainties shown include the statistical uncertainty, and a 15\% systematic uncertainty from the simulated HLT to account for the differences between our simulation of the trigger and the full experimental treatment.

As can be seen, the MPC displaced vertex technique constitutes a massive increase in trigger efficiency over the status quo. This increase in efficiency allows us to trigger on events where a Higgs produces displaced jets which currently easily evade detection. The displaced jet trigger is the single HLT most sensitive to these $h\to XX \to 4b$ events (though a combination of all HLTs could reach efficiencies as high as $\sim 5\%$, combining the data set in this way is prohibitively impractical). Many events pass the L1 trigger, but in a standard analysis the requirement to pass the HLT reduces the efficiency by nearly a factor of 50. The fast processing time of the MPC trigger allows us to place it before the HLT, and thus we do not suffer the efficiency penalty from satisfying the stricter requirements of the higher-level trigger.

We now broaden the class of models by considering the production of a second, non-SM Higgs $h'$, decaying pairs of $X$ scalars which in turn decay to $b\bar{b}$ with a displaced $c\tau$. The production cross section of this $h'$ we take as a free parameter. We consider a slice of the three-dimensional parameter space of $m_{h'}$, $m_X$, and $c\tau$ by setting $m_X = 20$~GeV. In Figure~\ref{fig:massctau}, we show the $h'$ production cross section times branching ratio to $XX$, necessary for a 10 event discovery after 1~fb$^{-1}$ of LHC14 data. As a reminder, the SM 125~GeV Higgs will have a production cross section of 49~pb at LHC14~\cite{Dittmaier:2011ti}. Thus, the MPC-driven trigger would be able to discover SM $h\to XX \to 4b$ decays with a branching ratio as small as  $6 \times 10^{-4}$ in a mere 1~fb$^{-1}$.

Using the presently existing single most sensitive HLT, a ten-event discovery scenario would require a BR~$>4-10\%$ for the 125~GeV Higgs, which rises quickly for a $h'$ lighter than this.  For comparison, the present CMS exclusion reach using $20$~fb$^{-1}$ of data from LHC8 requires $\sigma \times\mbox{BR} \gtrsim 10$~fb for $c\tau < 100$~mm with a 400~GeV $h'$ and $m_X = 50$~GeV. The luminosity requirements are at least an order of magnitude higher for a 200~GeV Higgs, and no bound is shown for a lighter Higgs \cite{Chatrchyan:2012cg}.  A combined LEPII analysis can place limits on $h' \to 4b$ below 125~GeV, but only excludes cross sections $\gtrsim 0.1\sigma_\text{SM}$ for masses above 80~GeV \cite{Schael:2006cr} (though this analysis does not require displaced decays). 

We should also emphasize that our HLT efficiencies are based on the current trigger menu in use at CMS. At the high luminosity LHC14, it is certain that the requirements for the HLT triggers will be raised, resulting in even lower efficiencies than presently indicated. While this statement applies equally to L1 thresholds, using a MPC-driven trigger side-steps the more severe thresholds that are inherent to the HLT.  Without a significant improvement of the trigger stream after the L1, such as our proposed MPC-based system, a significant range of displaced Higgs parameter space would remain out of reach.

Displaced decays involving new scalars and either the 125~GeV Higgs or as-yet-undiscovered additional doublet Higgses can be trivially implemented in simple extensions of the Higgs sector. Such scenarios are not well-constrained by existing LHC, Tevatron, or LEPII collider searches. Assuming decays through a long-lived $X$ into $b\bar{b}$ pairs, the existing search strategies at ATLAS and CMS are hampered by the low efficiency of the displaced $b$-jets to pass the HLT triggers. This is most notable for Higgs masses below $\sim 200$~GeV, as can be seen in Figure~\ref{fig:125higgsctautrigger}.

At best, under the current trigger configuration large amounts of luminosity could be required for a discovery in this mass region. To search for these models, novel techniques are required. A new set of triggers, based on fast Hough transforms performed by massively parallel computing \cite{Halyo:2013iba,Halyo:2013gja,Halyo:2013cza,Halyo:2013yfa}, would allow events to be triggered on based on global characteristics -- such as the existence of displaced vertices. Using this type of trigger, displaced decays of the Higgs could be discovered in early LHC14 running, even with a branching ratio as small as $\sim 10^{-3}$.

\end{document}